\documentclass[twocolumn,pre,showpacs,preprintnumbers,amsmath,amssymb]{revtex4}

\usepackage{graphicx}
\usepackage{dcolumn}
\usepackage{bm}
\usepackage{epsf}
\usepackage{mathrsfs}

\begin{document}
\title{Optically induced orientational transitions in nematics with planar alignment}

\author{Dmitry O. Krimer}
\email{krimer@pks.mpg.de}
\affiliation{Max Planck Institute for the Physics of Complex
Systems, N\"othnitzer Str. 38, D-01187 Dresden, Germany}

\date{\today}

\begin{abstract}
 Theoretical study of dynamical phenomena induced by a linearly polarized plane wave incident perpendicularly on a planar aligned nematic layer with the light intensity as the control parameter is reported. We find the threshold of the Optically Induced Fr\'eedericksz Transition for the planar state as a function of the problem parameters. It occurs that the threshold is substantially lower than that expected before. Above the primary instability the director settles either to a stationary or to an oscillatory states depending on a thickness of the layer. These states become unstable at a secondary threshold through a heteroclinic bifurcation and the director settles to a new stationary distorted state.
\end{abstract}

 \maketitle


Optical phenomena exhibited by nematic liquid crystals has been a subject of intensive study during the last few decades. A nematic
behaves optically as a uniaxial anisotropic medium with the
optical axis along the local molecular orientation described by
the director ${\bf n(r},t)$. The dielectric tensor governing the
propagation of light is anisotropic and depends on ${\bf n}$.
There are two competing mechanisms which determine the nematic alignment. On one hand, it is enforced by the anchoring forces at the confining surfaces. On the other hand the electric field of the incident light exerts a torque which may conflict with the boundary-imposed alignment. The torque increases with an increase in the light intensity. Eventually, at a certain light intensity the boundary-imposed alignment becomes unstable being replaced by the light-imposed one (Optically Induced Fr\'eedericksz Transition) which in turn affects the light propagation \cite{Tabiryan86_review}. A large variety of nonlinear phenomena occurs as a result this feedback
\cite{krimer_rev}. The possibility of dynamical changes of the nematic refractive index employing the orientational phenomena, has
attracted recently much attention in the context of construction
of all-optical devices based on photonic structures infiltrated
with liquid crystals, see e.g. \cite{Mirosh_08_1} and references therein. The goal of this paper is elaborated theoretical study of light induced phenomena in a nematic cell with planar alignment (without photonic structure) which is much less explored than that with the homeotropic one. This problem was considered before in
\cite{Santamato} where a linear analysis of the planar state with respect to a restricted class of perturbations was done. In the present paper we perform the general linear analysis with respect to
arbitrary (small) perturbations. We show that the planar state
might lose stability either through a stationary bifurcation or through a Hopf bifurcation when the thickness of the layer is above a certain
critical value. The thresholds for the primary instability are
substantially lower than those predicted before. Above the Hopf
bifurcation the system settles to an oscillatory state. This
state, in turn, becomes unstable at some higher value of intensity
where the transition to a new stationary distorted state via a
heteroclinic bifurcation occurs.

%
\begin{figure}
\vspace*{-2mm}
\begin{center}
\rotatebox{0}{\includegraphics[width=6cm]{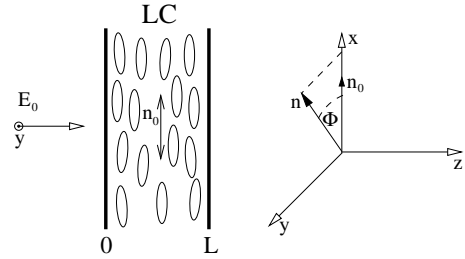}}
\end{center}
\vspace*{-4mm} \caption{Geometry of the setup: linearly polarized
light along the ${\bf y}$ direction incident perpendicularly on a
nematic layer with the director ${\bf n_0}
\parallel {\bf x} $ (planar state). The components of the
director ${\bf n}$ are described in terms of the twist angle
$\Phi$ ($\Phi=0$ in the planar state).} \label{fig1}
\vspace*{-6mm}
\end{figure}
%

We consider a linearly polarized along the $y$ direction plane wave, propagating along the $z$ axis. The wave incidents perpendicularly to a nematic layer, see Fig.~\ref{fig1}. Initially the director is orientated along the $x$ axis (planar alignment) and keeps its orientation at the boundaries (rigid boundary conditions). We restrict the consideration analyzing only spatially-homogeneous in the layer plane solutions which depend on the $z$  coordinate solely. To describe the director orientation  the twist angle $\Phi(z,t)$ is introduced, so that ${\bf n}=(\cos \Phi,\sin \Phi,0)$. The starting point is the set of nematodynamic equations coupled with Maxwell's equations for the propagation of light \cite{Gen}. The dynamical equation of motion for $\Phi(z,t)$ is obtained from the balance of torques (elastic, electromagnetic and viscous) acting on the nematic. The components of the electric field might be represented in terms of the amplitudes of the ordinary and extraordinary waves $A_e,A_o$ that vary slowly with $z$ on the scale $(k_0L)^{-1}$ (see e.g. \cite{disser}), where $k_0=\omega/c$ is the wavenumber of the incident wave and $L$ stands for the layer thickness. Finally, in the absence of a velocity field the coupled PDE for $\Phi$ and ordinary differential equations (ODEs) for $A_e$ and $A_o$ can be written as follows \cite{disser}
\begin{eqnarray}
\label{eq_phi_fin} \!\!\!\!\!&&\partial_t \Phi=\partial_z^2\Phi+
2\rho\,\tilde k_0^2 \,Re \left[A_e A_o^{\star} {\bf e}^{i \tilde
k_0 z } \right]
\\
\label{sys_Aoe_fin_1} \!\!\!\!\!&&\partial_z A_o=-(\partial_z
\Phi) {\bf e}^{i \tilde k_0 z } A_e,\,
\partial_z A_e=(\partial_z \Phi) {\bf e}^{-i \tilde k_0 z} A_o,
\end{eqnarray}
where $\tilde k_0$ is dimensionless thickness $\tilde k_0=2L\delta
n/\lambda$  with $\delta n=n_e-n_o$ ($n_e$ and $n_o$ are,
the refractive indices of the ordinary and extraordinary
light, respectively) and $\lambda$ is the incident light wavelength. To make the variables in Eqs.~(\ref{eq_phi_fin}), (\ref{sys_Aoe_fin_1}) dimensionless they have undergone the following scale transformation: $t\rightarrow t/\tau$;  $z\rightarrow \pi z/L$; amplitudes $A_{e,o}$ have been normalized over the electric field amplitude of the incident light; $\rho=I/I_c,$ where $I$ stands for the incident light intensity. Here $\tau=\gamma_1 L^2/\pi^2 K_2$ is characteristic relaxation time of the director motion and $I_c=8\pi^2 c K_2n_e\delta n/\lambda^2(n_e+n_o)$, $K_2$ is the twist elastic constant and
$\gamma_1=\alpha_3-\alpha_2$ is the rotational viscosity.

The boundary condition for $\Phi$ and initial conditions for
$A_o$, $A_e$ at $z=0$  read as follows:
\begin{eqnarray}
\label{bound_cond} \Phi_{z=0,\pi}(t)=0,\,\,|A_{o0}|^2=1,\,\,
|A_{e0}|^2=0,\,\,A_{e0}A_{o0}^{\star}=0.
\end{eqnarray}
Note that owing to the reflection symmetry with respect to the $y$ direction Eqs.~(\ref{eq_phi_fin}-\ref{bound_cond}) are invariant under the transformation S:$\left\{\Phi,A_e,A_o\right\}$ $\rightarrow$
$\left\{-\Phi,\mp A_e,\pm A_o \right\}.$ It is convenient to represent the solution of Eqs.~(\ref{eq_phi_fin})-(\ref{bound_cond}) as a series $\Phi(z,t)=\sum_n\Phi_n(t)\sin(nz)$, where each term of the sum satisfies the boundary conditions Eq.~(\ref{bound_cond}) identically. Then, Eqs. (\ref{eq_phi_fin})-(\ref{bound_cond}) are transformed into an infinite set of coupled equations for $\Phi_n$. To make the problem tractable the set is truncated at a certain large enough number of equations $N$ which are solved numerically by the   standard Runge-Kutta method. The error caused by the truncation is controlled by test runs with double and triple number of the modes. For every set of parameters $N$ is selected so that the difference between the routine and test runs is better than 1\%. Regarding equations for
$A_o,A_e$, they have to be solved dynamically at each step of numerical integration for time $t$. In addition, we perform a linear
analysis (numerically) of the stationary distorted states
($\partial_t \Phi_n=0$) by calculating the corresponding
Jacobians.

The starting point of the study is the linear stability analysis of the planar state ($\Phi(z,t)=0$). The linearized integro-differential
equation for $\Phi(z,t)=\Phi(z)\exp(\sigma t)$ is as follows:
\begin{eqnarray}
\label{my_lin} \partial_z^2 \Phi\! +\!2 \rho \tilde k_0^2 \left(
\Phi\!+\!\tilde k_0\int_0^z\Phi(z')\sin[\tilde k_0(z'-z)]
dz'\right)\!=\! \sigma\Phi.
\end{eqnarray}
It results in the following set of equations for $\Phi_n$ (eigenvalue problem):
\begin{eqnarray} \label{eigen_val_my}
&&\!\!\!\!\!\!\!\!\sum_n A_{mn}\Phi_n=\sigma \Phi_m
\\\nonumber
\label{Amn} &&\!\!\!\!\!\!\!\! A_{mn}\!=\!\left(\dfrac{2\rho
\tilde k_0^2}{n^2-\tilde
k_0^2}\!-\!1\right)n^2\delta_{mn}\!+\!\dfrac{4(-1)^m \rho \tilde
k_0^3mn\sin(\pi\tilde k_0)} {\pi(m^2-\tilde k_0^2)(n^2-\tilde
k_0^2)}.
\end{eqnarray}
The solvability condition for Eq.~(\ref{eigen_val_my}) requires det(A $-\; \sigma I$) = 0. It brings about infinite number of eigenvalues $\sigma_n$. However, at $n \gg \tilde k_0^2$ or (and) $m \gg \tilde k_0^2$ the off diagonal elements of matrix $A_{mn}$ decay as $1/n$ or (and) $1/m$, while the diagonal elements equal $-n^2$. Then, in the leading approximation det(A $-\; \sigma I$) factorizes for a product of infinite number of the diagonal elements at $n > N$ (which give stable real eigenvalues $\sigma_n \approx -n^2$ related to elastic relaxation) and a minor $N\times N$, where $N \gg \tilde k_0^2$. Eigenvalues of the minor are related to light-induced perturbations and should be inspected more carefully. These arguments provides us with the natural truncation scale $\sim \tilde k_0^2$. Note that the minimal number of modes for adequate description of the phenomenon increases as $L^2$. The planar state looses stability when the real part of (at least) one of the eigenvalues $Re(\sigma_n)$ becomes positive. It should be stressed that matrix $A_{mn}$ is not Hermitian and may have complex eigenvalues, in other words unstable modes may be oscillatory.
\begin{figure}
\begin{center}
\rotatebox{0}{\includegraphics[width=6cm]{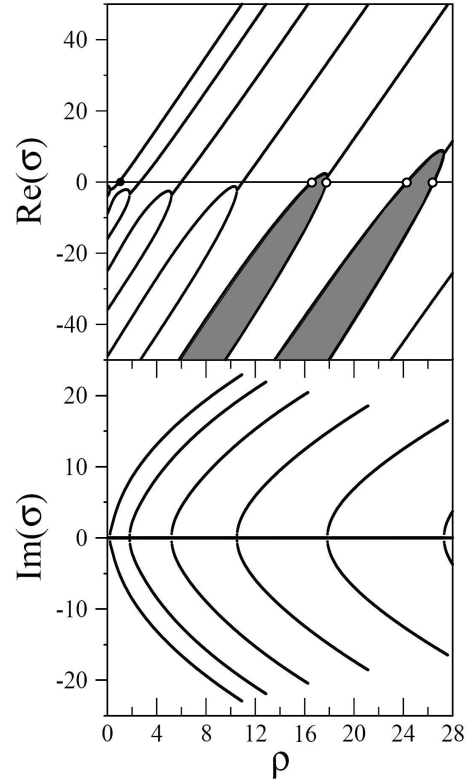}}
\end{center}
\vspace*{-6mm} \caption{$Re(\sigma)$ and $Im(\sigma)$ vs. $\rho$
for $\tilde k_0=1.6$. Filled circle: primary threshold
($\rho_1=0.95$, Hopf bifurcation). Shaded tongues cross abscissa
at, respectively, $\rho_1^\star=16.6$, $\rho_2^\star=17.8$,
$\rho_3^\star=24.3$ and $\rho_4^\star=26.5$ depicted by empty
circles.} \label{fig2} \vspace*{-6mm}
\end{figure}

\begin{figure}
\begin{center}
\rotatebox{0}{\includegraphics[width=6cm]{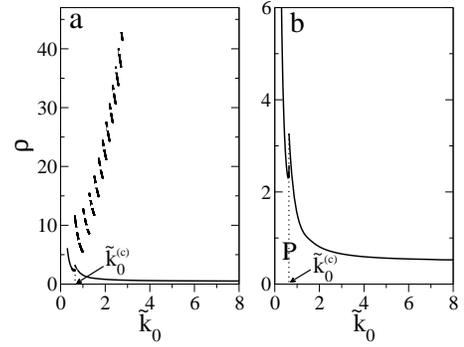}}
\end{center}
\vspace*{-5mm} \caption{Solid lines: stability diagram of the
planar state on the $(\tilde k_0,\rho)$ plane. Dashed lines on (a): thresholds calculated using the Eq.~(\ref{th_Sant}). Solid line
with $\tilde k_0<\tilde  k_0^{(c)}$ ($\tilde k_0 \ge \tilde
k_0^{(c)}$) corresponds to stationary (Hopf) bifurcation. P:
region of stability of the planar state.} \label{fig3}
\end{figure}

Typical results of the discussed stability analysis are presented on Fig.~\ref{fig2}, where $Re(\sigma_n)$ and $Im(\sigma_n)$ versus $\rho$ are shown for $\tilde k_0=1.6$. (The calculation  are made for $N=20$.) It is seen that $Re(\sigma_n)$ forms a tongued structure with a family of branches which, at first, appear as pairs of purely real branches that go almost parallel to each other as $\rho$ increases, but then the pair merges at a certain $\rho\!\ge\! \rho_{top}^{(n)}$. Next, the branches continue as a pair of complex conjugate modes. Some of the purely real branches cross the zero line at intensities $\rho$ given by the formula
\vspace*{-2mm}
\begin{eqnarray}
 \label{th_Sant} \dfrac{2\rho  \cdot \sin[\pi\tilde
k_0 \sqrt{1+2\rho}]}{\pi\tilde k_0 \sqrt{1+2\rho}}=-1,
\end{eqnarray}
which was derived in \cite{Santamato}. The fist two tongues
depicted by shaded areas crosses the abscissa at,
$\rho_1^\star=16.6$, $\rho_2^\star=17.8$, $\rho_3^\star=24.3$ and
$\rho_4^\star=26.5,$ respectively, and are the same as ones shown in Fig.~1 of \cite{Santamato}. (The dimensionless $\tilde L=5$ introduced there corresponds to $\tilde k_0=\tilde L/\pi\sim 1.6$ used in our
calculations.) In \cite{Santamato} such a structure was interpreted as a series of alternate stable and unstable intervals for the planar state as the light intensity increases with the lowest threshold for instability given by $\rho_1^\star=16.6$. However, in reality, the planar state looses stability via a Hopf bifurcation at much lower value $\rho_1=0.95$ (filled circle on Fig.~\ref{fig2}) and {\it never} restores stability at higher intensities again. In fact, the reason for such a difference with the results of \cite{Santamato} is that for solving the eigenvalue problem (5) therein a particular (not a general) form of a trial solution was used [see \cite{Santamato}, Eqs.~(6,10)]. As a result, the growth rates turned out to be {\it a priory} real instead of being complex and, thus, the ones with $Im(\sigma)\neq 0$ predicted here have been not caught. The stability diagram in the $(\tilde k_0,\rho)$ plane is shown in Fig.~\ref{fig3} by solid lines and the thresholds calculated using the Eq.~(\ref{th_Sant}) are depicted by dashed lines. It is interesting, that below a certain critical value $\tilde k_0^{(c)}<0.64$ (corresponding to $\tilde L=2$ in \cite{Santamato}), the primary bifurcation is
stationary indeed and the values of thresholds are correctly described by Eq.~(\ref{th_Sant}). In that case, the very first tongue
crosses the $\rho$ axis (in contrast to the situation for $\tilde
k_0\!=\!1.6\!>\!\tilde k_0^{(c)}$ depicted in Fig.~\ref{fig2}
where the first tongue is under the abscissa) and at the threshold
the largest growth rate corresponds to the purely real branch. When approaching $\tilde k_0^{(c)}$ from below, this first tongue goes down and, finally, at $k_0=\tilde k_0^{(c)}$ the whole tongue lies under the $\rho$ axis. Note, that transition to another branch with $Im[\sigma]\neq 0$ is accompanied by a stepwise change of the threshold intensity. Such a behavior is related to the inclined structure of
the tongues (see Fig.~\ref{fig3}). It should be stressed that for $\tilde k_0>\tilde k_0^{(c)}$, the true values of the thresholds differs from the ones predicted by Eq.~(\ref{th_Sant}) in one order, and with increase of $\tilde k_0$ quite rapidly, in two orders of
magnitude. The true threshold of the primary instability, decreases with an increase in $\tilde k_0$ from $I=3.3\, I_c$ at $\tilde k_0=\tilde k_0^{(c)}$ to $0.5\,I_c$ at $\tilde k_0=8$ and then practically does not change at further increase in the $\tilde k_0$, see Fig.~\ref{fig3}. To obtain a stability diagram in the dimensional $(L,I)$ plane one should rescale $\tilde k_0$ and $\rho$-axes of Fig.~\ref{fig3} multiplying them by $\lambda/(2\,\delta n)$ and $I_c$, respectively. In experiment one should take nematics with rather small values of $\delta n$ to avoid very large values for thresholds which are of the order of $I_c$. For instance, for $\delta n=5\cdot
10^{-3}$, $K_2=2.5 \cdot 10^{-7}\,dyn$ and $\lambda=532\, nm$ the
value of $I_c$ is $52\,kW/cm^2$ and the critical length
$L^{(c)}=\tilde k_0^{(c)} \lambda/(2\,\delta n)$ is $34\,\mu m$.

%
%
\begin{figure}
\begin{center}
\rotatebox{0}{\includegraphics[width=6cm]{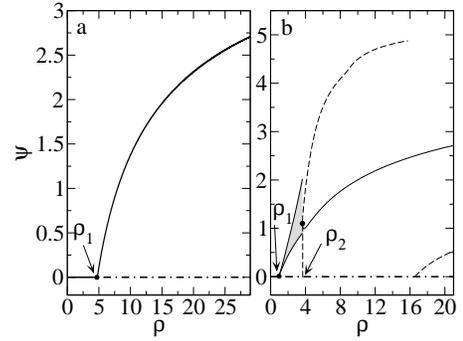}}
\end{center}
\vspace*{-5mm} \caption{Bifurcation diagrams for (a) $\tilde
k_0=0.5$ and (b) $\tilde k_0=1.6$: parameter of reorientation
$\Psi$ vs. $\rho$. Solid (dashed) curves correspond to stable
(unstable) stationary solutions. $\rho_1$: primary instability of
the planar state. (a) $\rho_1=4.76$ (stationary bifurcation); (b)
$\rho_1=0.95$ (Hopf bifurcation). Gray region (confined by two
lines): oscillatory states of the director. $\rho_2=3.65$:
secondary instability (heteroclinic bifurcation). } \label{fig4}
\vspace*{-4mm}
\end{figure}
%
It is worth noting, that the linear integro-differential equation
(\ref{eigen_val_my}) appears in the framework of linear analysis
of another problem, namely the one with a linearly polarized ordinary
light wave incident at a small oblique angle on a thin layer of
homeotropically oriented nematic \cite{Tab_98,demeter00}. The only
difference is that there one deals with the normalized incidence
angle $\kappa$ instead of $\tilde k_0$ used here. It is well known
that stability diagram in the $(\rho,\kappa)$ plane might be
calculated with good accuracy in the framework of a two modes
approximation (for $\kappa<1$) and consists of a line of
stationary instability which joins the line of Hopf instability in
the Takens-Bogdanov point \cite{demeter00}. The line of stationary
instability exists only below a certain critical value of $\kappa^{(c)}$ and as we checked out is indeed described by the formula (\ref{th_Sant}) after rewriting it in corresponding quantities.
However, above $\kappa^{(c)}$ the homeotropic state looses
stability always via a Hopf bifurcation and the formula
(\ref{th_Sant}) is not applicable anymore. Moreover, the
homeotropic state will never be alternatively stable and unstable
for any $\kappa$ and at higher intensities a very complex dynamics
occurs \cite{krimer_rev}. Here we have a similar picture but with
respect to the critical $\tilde k_0^{(c)}$.

We choose a sum of squares of all modes $\Psi=\sum_n \Phi_n^2$ as
a theoretical measure of reorientation inside nematic. It is an
appropriate quantity because $\Psi=0$ for the planar state, it
increases with increasing of the reorientation, saturates to a
certain value since the amplitudes damp after certain $n$ and is
independent on material parameters. In Fig.~\ref{fig4} typical
bifurcation diagrams are shown for $\tilde k_0<\tilde k_0^{(c)}$
and $\tilde k_0>\tilde k_0^{(c)}$. The planar state remains stable
when $\rho<\rho_1$. At $\rho=\rho_1$ we deal with a continuous
transition via either a stationary for $\tilde k_0<\tilde
k_0^{(c)}$ or a Hopf bifurcation in opposite case. Above the
threshold and for $\tilde k_0<\tilde k_0^{(c)}$ the director
settles to a stationary distorted state, whereas for $\tilde k_0>
\tilde k_0^{(c)}$ to an oscillatory one.  In latter regime, the
lower and the upper lines depicted in Fig.~\ref{fig4}(b) bound the
region in gray and correspond to the minimum and maximum values
taken by $\Psi$ during its oscillation. The director motions
develops along the limit cycle in the space of $\Phi_n$ (which is
infinite). Importantly, the planar state is always unstable for
$\rho>\rho_1$ in both cases.
%
%
\begin{figure}
\begin{center}
\rotatebox{0}{\includegraphics[width=5.5cm]{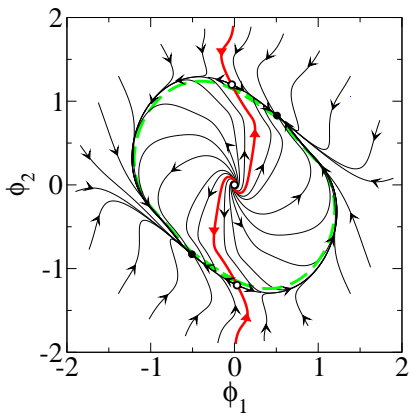}}
\end{center}
\vspace*{-7mm} \caption{Dynamics near the secondary threshold
$\rho_2=3.6530$: director trajectories at $\rho=3.7967>\rho_2$ in
the plane of first two modes $(\Phi_1,\Phi_2)$. Starting from
different initial condition the system settles to one of the two
stable nodes (filled circles). Red curve: separatrix. Empty
circles: unstable fixed points (saddles). Origin: unstable spatially uniform planar state (focus). Dashed green line: limit cycle at
$\rho=3.638<\rho_2$.} \label{fig5} \vspace*{-5mm}
\end{figure}
%
%
In some narrow region around $\rho_2\!=\!3.65$ the period of oscillations increases progressively with increasing light intensity, and oscillations become substantially anharmonic. The period diverges at $\rho=\rho_2$ which correspond to a secondary bifurcation transforming the system into a new stationary distorted state. The dynamics near the threshold is summarized in the projection of the true phase trajectory on the plane $(\Phi_1,\Phi_2)$, see Fig.~\ref{fig5}. The bifurcation at $\rho=\rho_2$ belongs to rather a rare type and corresponds to the following. Above $\rho_2$ a system has two pairs of stationary nontrivial solutions which are mutual images under the symmetry transformation $S$. The first pair is represented by a stable node and its image and the second one by a saddle and its image. There is also a trivial solution which is represented by an unstable focus corresponding to the spatially uniform planar state. Starting from different initial conditions the system eventually settles to one of the stable nodes. A separatrix, which partitions phase plane onto the basins of attraction for the two stable nodes, goes through the two saddles and the unstable focus. Approaching $\rho_2$ from above, the saddle and node go closer and closer to each other and finally merge at $\rho=\rho_2$. At this point the limit cycle appears which exist for $\rho_1<\rho<\rho_2$.  It is worth noting that the symmetry $S$
is spontaneously broken at the secondary bifurcation. We checked
that the bifurcation scenario displayed in Fig.~\ref{fig4} exists
over wide region of $\tilde k_0$. However, the oscillatory regime
seems to be more complex for $\tilde k_0\ge 3$ than that described above.

In conclusion, we have studied theoretically the transitions
induced by linearly polarized light incident perpendicularly to a
layer of nematic that has initially spatially uniform planar alignment. We have found the primary threshold as a function of (normalized)
thickness of the layer by performing a linear analysis of the
basic state. Our numerical analysis of the problem shows that,
with increasing light intensity, the planar state becomes unstable
in a favor of either a stationary distorted state when the
thickness is below a certain critical value, or to an oscillatory state if the thickness is above the critical value. As the intensity increases further, the oscillatory state disappears via a secondary bifurcation. At the bifurcation point two identical saddle-node fixed points are born on opposite sites of the limit cycle. It results in divergence of the period of oscillations and partition of the limit cycle into two heteroclinic orbits. Further increase in the intensity brings about splitting each of the fixed point into a stable node and a saddle. In typical cases the calculated thresholds have values substantially lower than those believed before.

The author wishes to thank M.I. Tribelsky, S. Flach, A.
Miroshnichenko, E. Brasselet and A. Krekhov for helpful discussions and comments.

\vspace*{-7mm}


\begin{thebibliography}{10}
%
\vspace*{-11mm}

\bibitem{Tabiryan86_review}
N.~V. Tabiryan, A.~V. Sukhov, and B.~Y. Zel'dovich,  Mol. Cryst.
Liq. Cryst. {\bf 136}, 1 (1986).

\bibitem{krimer_rev} G. Demeter and D.O. Krimer, Phys. Reports, {\bf 448}, 133 (2007)

\bibitem{Mirosh_08_1} A.~E. Miroshnichenko, E. Brasselet, Y.~S. Kivshar, Appl. Phys. Lett. {\bf 92} 53306 (2008); U.~A. Laudyn, A.~E. Miroshnichenko, W. Krolikowski, D.~F. Chen,
Y.~S. Kivshar, M.~A. Karpierz, Appl. Phys. Lett. {\bf 92} 203304
(2008)

\bibitem{Santamato} E.  Santamato, G. Abbater, P. Maddalena and Y.R. Shen,
Phys. Rev. A {\bf 36}, 2389 (1987).

\bibitem{Gen} P. G. de Gennes and J. Prost, The physics of liquid crystals (Clarendon press, Oxford, 1993).

\bibitem{disser}
D.~O. Krimer, PhD Thesis, University of Bayreuth (2004); URL:\\
http://opus.ub.uni-bayreuth.de/volltexte/2004/98/;  E. Brasselet, T.V. Galstian, L. J. Dube, D. O. Krimer and L. Kramer, J. Opt. Soc. Am. B, {\bf 22}, 1671 (2005).

\bibitem{Tab_98} N.V. Tabiryan, A.L. Tabiryan-Murazyan,
V. Carbone, G. Cipparrone, C. Umeton, C. Versace, T. Tschudi,
Optics Comm. {\bf 154}, 70 (1998).

\bibitem{demeter00} G. Demeter, Phys. Rev. E {\bf 61}, 6678 (2000);
D. O. Krimer, G. Demeter and L. Kramer, Phys. Rev. E {\bf 66},
031707 (2002).
%


\end{thebibliography}
\end{document}